\documentclass[prb,twocolumn,superscriptaddress,showpacs]{revtex4-1}
\usepackage{amsmath}
\usepackage{color}
\usepackage{hyperref}
\usepackage{graphicx}
\usepackage{txfonts}
\usepackage{feynmp}
\begin{document}
\title{Reconstructed Fermi surface and quantum oscillation of doped resonating valence bond state with incommensurate charge order in underdoped cuprates}
\author{Long Zhang}
\affiliation{Institute for Advanced Study, Tsinghua University, Beijing, 100084, China}
\author{Jia-Wei Mei}
\affiliation{Perimeter Institute for Theoretical Physics, Waterloo, Ontario, N2L 2Y5 Canada}
\date{\today}
\begin{abstract}
Recent experiments have revealed incommensurate charge density wave (CDW) in the pseudogap regime in underdoped cuprates, e.g. YBa$_2$Cu$_3$O$_{6+\delta}$ and HgBa$_2$CuO$_{4+\delta}$. However, its relationship with the pseudogap is still controversial. In this work, we take a phenomenological synthesis of the doped resonating valence bond (RVB) state and the CDW order. Starting from the Yang-Rice-Zhang Green's function ansatz for the doped RVB state [Phys. Rev. B {\bf 73}, 174501 (2006)], in which the Fermi surface is partially truncated into four nodal hole-like Fermi pockets by the antinodal RVB gap, we show that the CDW order at the wavevectors connecting the tips of the Fermi arcs (the hotspots) induces Fermi surface reconstruction, giving rise to an electron-like Fermi pocket ($\alpha$ orbit) and a new hole-like Fermi pocket ($\beta$ orbit). The $\alpha$ orbit is formed by joining the Fermi arcs at the hotspots and it dominates the quantum oscillation Fourier spectrum, while the $\beta$ orbit is formed by joining the outer patches of the original hole pockets, which has vanishingly small spectral weight. The areas enclosed by these orbits are extracted from the density of states oscillation in magnetic field and quantitatively agree with the experiments.
\end{abstract}
\maketitle

\section{Introduction}

Despite decades of intensive research, the origin of the pseudogap in the underdoped cuprates remains much debated. The pseudogap is characterized by the loss of low-energy density of states (DoS) as observed in the magnetic susceptibility, the specific heat and the transport measurements \cite{Timusk1999} and the antinodal gap in the angle-resolved photoemission spectroscopy (ARPES) \cite{Damascelli2003} below a doping-dependent pseudogap temperature. Recently, it is revealed that incommensurate charge density wave (CDW) shows up in the pseudogap regime of the clean YBa$_{2}$Cu$_{3}$O$_{6+\delta}$ (Y123) and HgBa$_{2}$CuO$_{4+\delta}$ (Hg1201) materials \cite{Wu2011b, Achkar2012, Chang2012, Ghiringhelli2012, LeBoeuf2012, Blackburn2013, Blanco-Canosa2013, Wu2013b, Shekhter2013, Blanco-Canosa2014, Huecker2014, Tabis2014, Wu2014, Comin2014, Comin2014b, Fujita2014}, which competes with the superconductivity \cite{Chang2012, Blackburn2013, Blanco-Canosa2013, Blanco-Canosa2014}. The charge order in the pseudogap phase has been extensively studied \cite{Kivelson1998, Yamase2000, Halboth2000, Chakravarty2001, Simon2002, Li2006a, Vojta2008, Metlitski2010, Vojta2012, Holder2012, Husemann2012, Bejas2012, Kee2013, Bulut2013, Efetov2013, Sachdev2013a, Hayward2014, Allais2014, Lee2014, Pepin2014}. In particular, the bidirectional CDW order can reconstruct the Fermi surface to form an electron pocket. This scenario has been adopted to explain \cite{Harrison2011, Allais2014, Seo2014} the negative Hall and Seebeck coefficients \cite{LeBoeuf2007, Chang2010, Doiron-Leyraud2013} and the quantum oscillation observed at low temperature and high magnetic field \cite{Doiron-Leyraud2007, Jaudet2008, Sebastian2008b, Audouard2009, Ramshaw2010, Sebastian2010, Sebastian2010b, Singleton2010, Laliberte2011, Riggs2011, Sebastian2011a, Sebastian2011, Sebastian2012a, Vignolle2013, Sebastian2014, Bangura2008, Yelland2008, Barisic2013}, which clearly demonstrates the presence of Fermi-liquid-like quasiparticles in this regime and triggers intensive research \cite{Chen2008, Chen2009a, Taillefer2009, Chubukov2010, Ma2013, Allais2014}.

However, the relevance of the CDW fluctuations to the \emph{origin} of the pseudogap remains controversial. The CDW onset temperature is lower than the pseudogap and the deviation is more significant for doping concentration $x<0.12$ \cite{Bakr2013, Blanco-Canosa2014, Huecker2014}. The sign change (from positive to negative) of the Hall and Seebeck coefficients occurs at even lower temperature \cite{LeBoeuf2007, Chang2010, Doiron-Leyraud2013}. Close to the optimal doping, the pseudogap develops, however, the CDW is not detected \cite{Mei2012a, Blanco-Canosa2014}. On the other hand, as already noted by Lee \cite{Lee2014}, the CDW induced gap on a large Fermi surface cannot fully account for the single-particle spectral feature in ARPES \cite{He2011}. Moreover, the CDW order in the La-based compounds, e.g., La$_{2-x}$Sr$_x$CuO$_4$ (LSCO) and La$_{2-x}$Ba$_x$CuO$_4$ (LBCO), is very different from the non-La-based compounds, e.g., Y123, YBa$_2$Cu$_4$O$_{8}$ (Y124) and Hg1201. Stripe order is generally found in the La-based compounds, whereas bidirectional CDW is found in the non-La-based compounds \cite{[For a review see ] Fradkin2014}. In this work, we focus on the non-La-based compounds. Despite the diversity of the CDW order forms in different cuprate families, the pseudogap behaviors are largely universal. It suggests that the CDW order cannot be taken as the driving force for the pseudogap phenomena; instead, it should be regarded as a secondary instability in this regime.

In this work, we treat the pseudogap and the CDW order as independent phenomena and provide a theoretical synthesis to show that the CDW order on top of a pseudogap state can capture the Fermi surface reconstruction and the doping evolution of the Fermi pocket areas measured in quantum oscillations in Y123, Y124 and Hg1201. We take the pseudogap state as a doped resonating valence bond (RVB) state with small nodal hole pockets, which is described by the Yang-Rice-Zhang (YRZ) ansatz of the electron Green's function \cite{Yang2006, Rice2012}. Other theoretical proposals for the truncated Fermi pockets in doped spin liquid are also plausible, e.g., the fractionalized Fermi liquid (FL$^{*}$) by Sachdev and collaborators \cite{Qi2010a, Moon2011} and the Luttinger-volume-violating Fermi liquid by Mei \textit{et al} \cite{Mei2012a, Mei2012}. A recent work has been carried out independently to study the CDW instability in the FL$^{*}$ state \cite{Chowdhury2014}. The YRZ Green's function reproduces the Fermi arcs at the Fermi energy observed by ARPES \cite{Damascelli2003} and a number of anomalous features in the optical spectroscopy and the thermodynamic measurements \cite{Rice2012}. In this paper, we try to find a compatible CDW order integrated on top of the YRZ hole-like Fermi pockets for the non-La-based cuprate compounds. It is established in experiments \cite{Comin2014b} that the CDW order occurs at the wavevectors connecting the tips of the Fermi arcs (the hotspots). Assuming that such a static incommensurate CDW order takes place on top of the YRZ state at low temperatures and high magnetic fields, we explicitly introduce the CDW order on top of the YRZ hole pockets and find that the Fermi surface is reconstructed and two magnetic orbits show up in the quantum oscillations. Different from the previous study \cite{Vojta2012}, where only hole pockets were found for the \emph{commensurate} CDW order on top of the YRZ state, we find an electron-like and a hole-like Fermi pockets due to the Fermi surface reconstruction.

\begin{figure}
\centering
\includegraphics[width=0.48\textwidth]{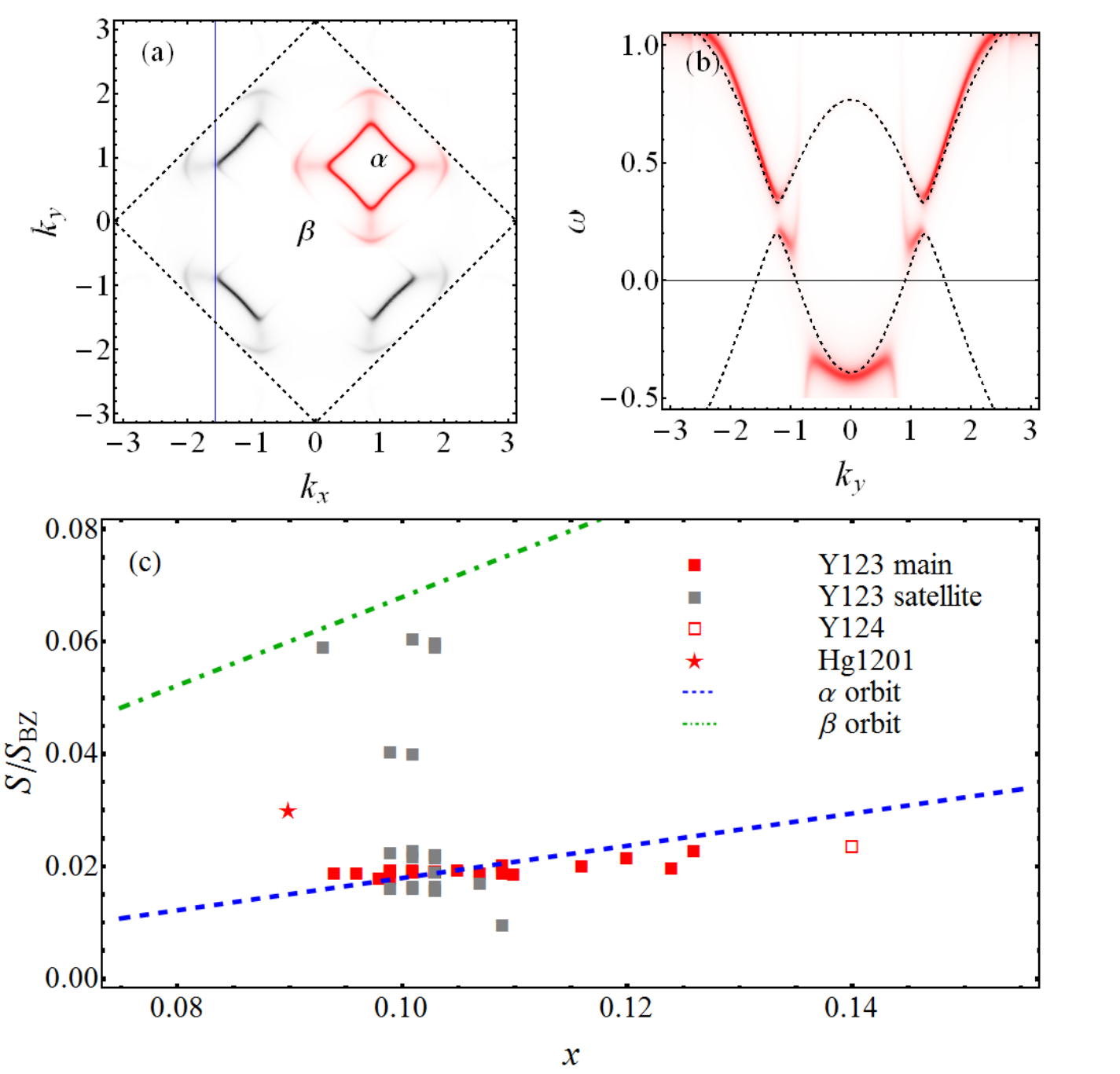}
\caption{(Color online) (a) Reconstructed Fermi surface due to the CDW order for doping $x=0.12$ and CDW order magnitude $P_{0}=0.3$. In the first quadrant of the Brillouin zone, the Fermi surface patches are joined up by shifting by the CDW wavevectors to illustrate the closed magnetic orbits: an electron-like $\alpha$ orbit and a hole-like $\beta$ orbit. (b) The energy distribution of the spectral function along the momentum cut shown in (a). The dashed curve is the energy dispersion of the YRZ state in the absence of the CDW order. A spectral gap opens at the Fermi energy. (c) The doping dependence of the reconstructed Fermi pocket areas. The dashed lines denote the $\alpha$ and $\beta$ orbits respectively. The results extracted from the quantum oscillation experiments \cite{Doiron-Leyraud2007, Jaudet2008, Sebastian2008b, Audouard2009, Ramshaw2010, Sebastian2010, Sebastian2010b, Singleton2010, Laliberte2011, Riggs2011, Sebastian2011a, Sebastian2011, Sebastian2012a, Vignolle2013, Sebastian2014, Bangura2008, Yelland2008, Barisic2013} are included for comparison.}
\label{FigQOvsDoping}
\end{figure}

Our main results are shown in Fig. \ref{FigQOvsDoping}. The static CDW order opens a spectral gap at the hotspots on the nodal hole pockets of the YRZ state shown in Fig. \ref{FigQOvsDoping} (b). The four Fermi arcs are joined up by shifting by the CDW wavevectors to form an electron-like Fermi pocket, denoted as the $\alpha$ orbit, while the other sides of the nodal hole pockets with vanishingly small spectral weight, the ``shadow'' patches, are also joined up to form a new hole-like Fermi pocket, denoted as the $\beta$ orbit shown in Fig. \ref{FigQOvsDoping} (a). Their areas satisfy $S_{\beta}-S_{\alpha}=4S_{\mathrm{YRZ}}=xS_{\mathrm{BZ}}/2=(4\pi^{2}/a_{0}^{2})x/2$, in which $S_{\mathrm{YRZ}}$ and $S_{\mathrm{BZ}}$ are the areas of one original nodal hole pocket and the entire first Brillouin zone respectively. $a_{0}$ is the lattice constant and $x$ is the hole doping concentration. These orbits can be clearly resolved in the calculated density of states (DoS) oscillation in magnetic field and the electron-like $\alpha$ orbit dominates the quantum oscillation, in agreement with experiments. The experiment results are collected in Fig. \ref{FigQOvsDoping} (c) for comparison. The doping dependence of the $\alpha$ orbit oscillation frequency quantitatively agrees with the dominant oscillation peak in experiments, and the $\beta$ orbit gives rise to the higher-frequency oscillation peak observed by Sebastian \textit{et al} \cite{Sebastian2008b, Sebastian2010, Sebastian2010b, Sebastian2011}. Although the existence of the $\beta$ orbit peak is controversial in experiments \cite{Audouard2009}, we suggest that this orbit can be taken as evidence of the ``shadow'' side of the nodal hole pockets.

The rest of this paper is organized as follows. The YRZ ansatz of the pseudogap state is briefly reviewed in Sec. \ref{SecYRZ}, in which we stress that the single-particle spectral features observed by ARPES are well reproduced. In Sec. \ref{SecFS}, the CDW wavevectors are identified by the local maxima in the CDW susceptibility and the static CDW order is introduced to study the Fermi surface reconstruction. The DoS oscillation in magnetic field is calculated in Sec. \ref{SecQO} and its robustness is verified. The main results are summarized in Sec. \ref{SecSum}.

\section{Yang-Rice-Zhang ansatz of pseudogap state} \label{SecYRZ}

\begin{figure}[b]
\centering
\includegraphics[width=0.48\textwidth]{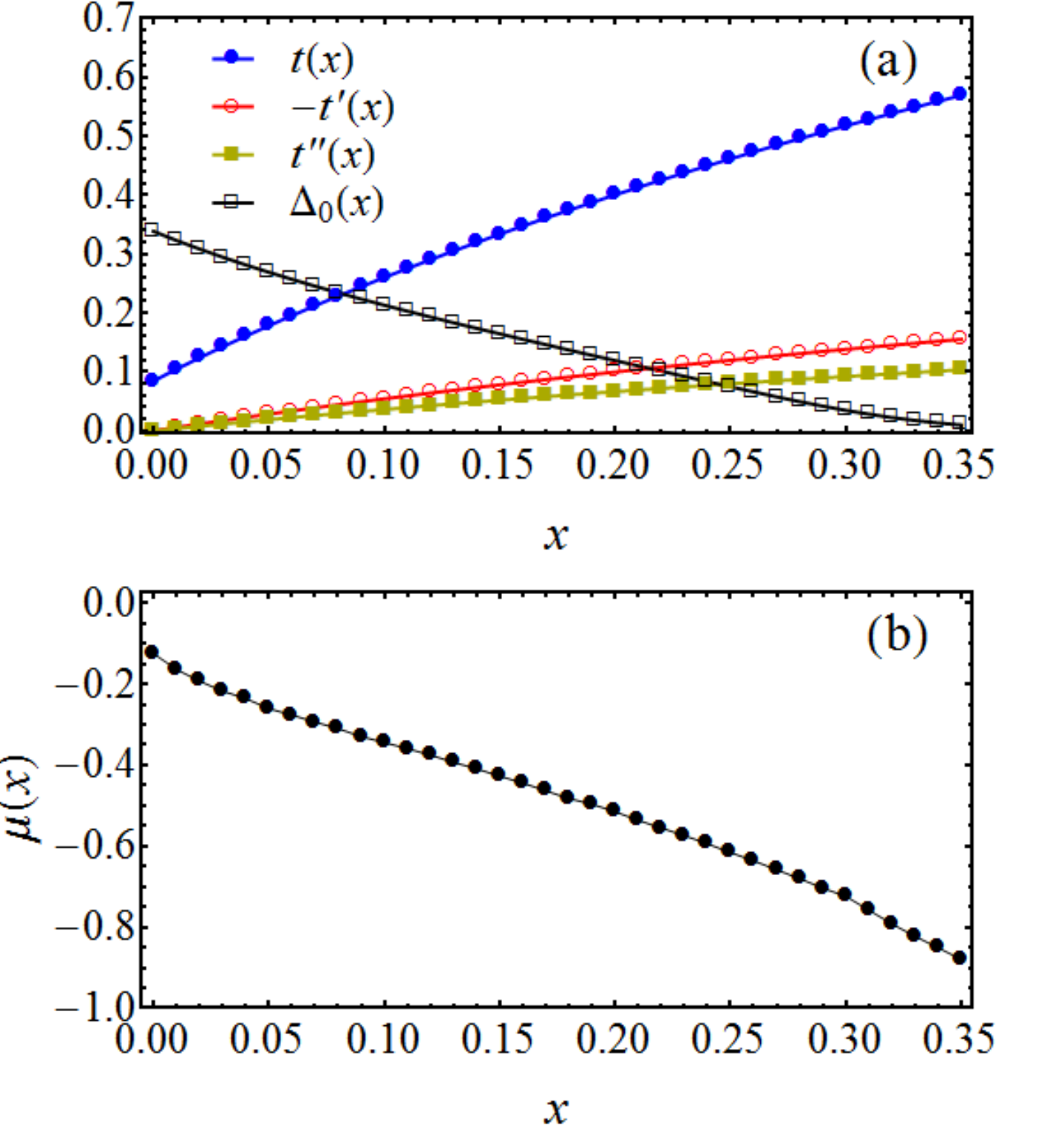}
\caption{(Color online) The doping dependence of (a) the renormalized hopping and RVB paring parameters $t(x)$'s and $\Delta_{0}(x)$ and (b) the chemical potential $\mu(x)$ adopted in the YRZ Green's function ansatz.}
\label{FigYRZparameters}
\end{figure}

We take the pseudogap phase as a doped RVB state with the coherent part of the electron Green's function described by the Yang-Rice-Zhang (YRZ) ansatz \cite{Yang2006, Rice2012},
\begin{equation} \label{EqYRZ}
G_{0}(\omega, \vec{k})=\frac{g_{t}(x)}{\omega-\xi(\vec{k})-\Sigma_{\mathrm{RVB}}(\omega, \vec{k})},
\end{equation}
in which $\xi(\vec{k})=-2t(x)(\cos k_{x}+\cos k_{y})-4t'(x)\cos k_{x}\cos k_{y}-2t''(x)(\cos 2k_{x}+\cos 2k_{y})-\mu(x)$ is the energy dispersion with up to the third-nearest-neighbor hopping terms. The self-energy ansatz $\Sigma_{\mathrm{RVB}}(\omega, \vec{k}) = \Delta(\vec{k})^{2}/(\omega +\xi_{0}(\vec{k}))$, with $\xi_{0}(\vec{k})=-2t(x)(\cos k_{x}+\cos k_{y})$ and the $d$-wave RVB pairing amplitude $\Delta(\vec{k})=\Delta_{0}(x)(\cos k_{x}-\cos k_{y})$. This self-energy ansatz was proposed by analogy with that of the doped spin liquid on a ladder \cite{Konik2006, Yang2006}. An alternative derivation for this form of self-energy based on the slave-boson theory is given in Ref. \onlinecite{James2012}. The hopping parameters $t(x)=g_{t}(x)t+3g_{J}(x)J\chi(x)/8$, $t'(x)=g_{t}(x)t'$ and $t''(x)=g_{t}(x)t''$ are renormalized from the bare band parameters \cite{Mattheiss1990, Yang2006} $t$, $t'=-0.3t$, $t''=0.2t$ and $J=t/3$ according to the renormalized mean field theory (RMFT) \cite{Zhang1988}, in which $g_{t}(x)=2x/(1+x)$ and $g_{J}(x)=4/(1+x)^{2}$ capture the impact of the single occupancy constraint in a doped Mott insulator (Gutzwiller approximation) and the mean field parameters $\chi(x)$ and $\Delta_{0}(x)$ are determined self-consistently by the RMFT \cite{Yang2006, Zhang1988}. The details of the RMFT calculations are summarized in Appendix \ref{AppRMFT}. The chemical potential $\mu(x)$ is adjusted to guarantee the generalized Luttinger theorem
\begin{equation}
\frac{2}{4\pi^{2}/a_{0}^{2}}\int_{G(0,\vec{k})>0}d^{2}\vec{k}=2-x.
\end{equation}
The doping dependence of these parameters is plotted in Fig. \ref{FigYRZparameters}.

\begin{figure}
\centering
\includegraphics[width=0.48\textwidth]{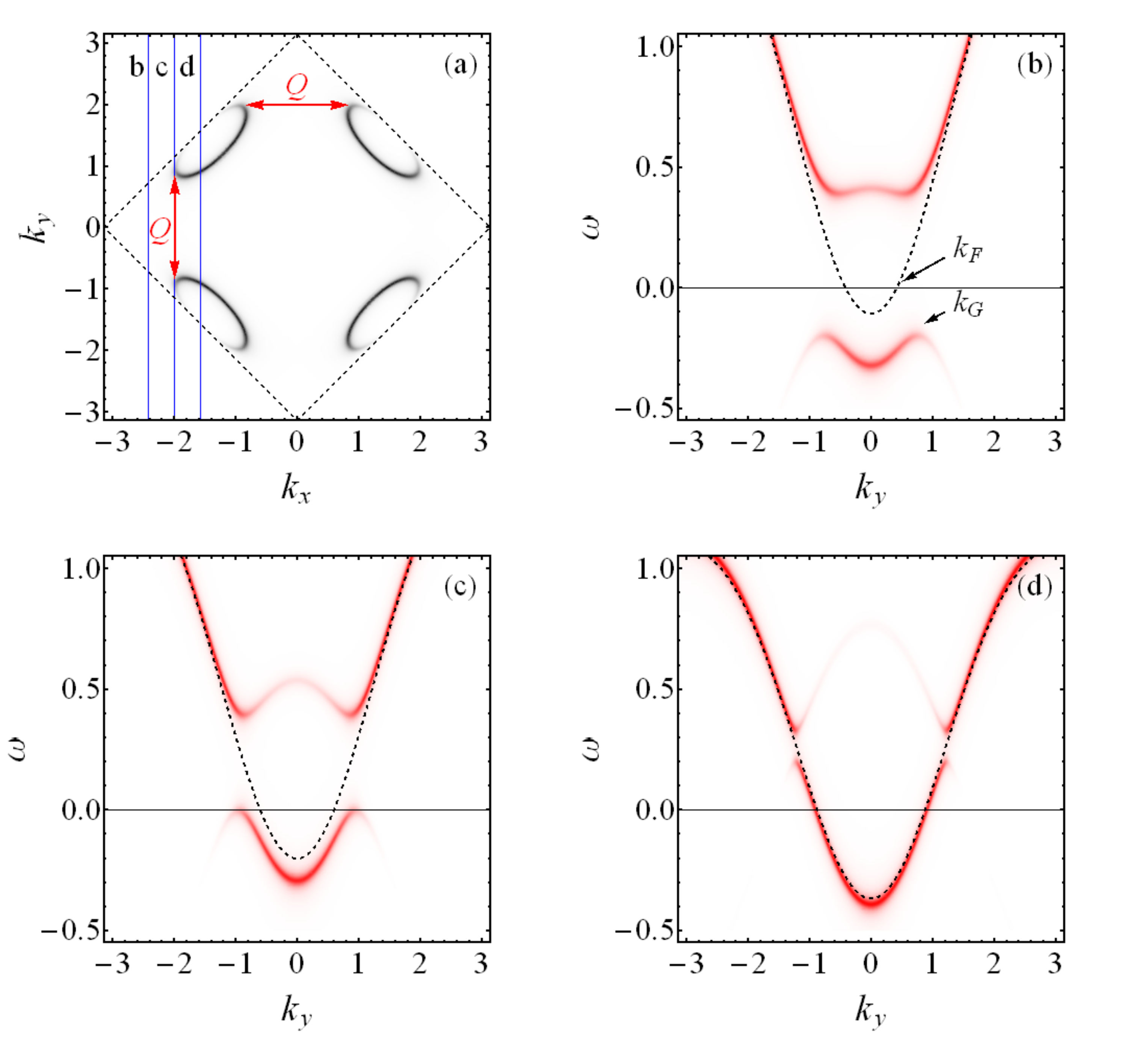}
\caption{(Color online) (a) The single-particle spectral function at the Fermi level $A(i\eta, \vec{k})$ of the YRZ Green's function with doping $x=0.12$, $\eta=0.0003t\simeq 1\mathrm{meV}$. The dashed lines indicate the magnetic Brillouin zone boundary $k_{x}\pm k_{y}=\pm \pi$, where the self-energy diverges. The arrows indicate the wavevectors connecting the hotspots. (b)--(d) The energy dependence of $A(\omega,\vec{k})$ along the momentum cuts shown in (a) across the antinodal region, the hotspots and the nodal region, respectively. The dashed curves show the dispersion of the normal state defined by setting $\Delta_{0}(x)=0$. $k_{\mathrm{G}}$ indicates the momentum where the lower energy band bends back and the minimal spectral gap opens, while $k_{\mathrm{F}}$ indicates the Fermi momentum of the normal state. They do not coincide with each other, as found in ARPES measurements \cite{Hashimoto2010, He2011, Yang2011}.}
\label{FigYRZSpectral}
\end{figure}

In Fig. \ref{FigYRZSpectral} (a), we show the spectral function $A(\omega, \vec{k})=-\pi^{-1} \mathrm{Im} G_{0}(\omega+i \eta, \vec{k})$ at the Fermi level $\omega=0$ for doping $x=0.12$. The Green's function poles form four hole pockets in the nodal region and the area of each pocket is $(4\pi^{2}/a_{0}^{2})x/8$. The spectral weight is vanishingly small on the outer sides of the pockets near the lines $k_{x}\pm k_{y}=\pm \pi$ due to the divergence of the self-energy $\Sigma_{\mathrm{RVB}}(0,\vec{k})$ at these lines, so the Fermi arc feature in ARPES is captured. In Figs. \ref{FigYRZSpectral} (b)--(d), we show the energy dependence of $A(\omega, \vec{k})$ along the momentum cuts in Fig. \ref{FigYRZSpectral} (a). In the antinodal region, the minimal spectral gap indicated by the backbending of the lower energy band ($k_{\mathrm{G}}$) does not open at the Fermi momentum ($k_{\mathrm{F}}$) of the normal state defined by setting $\Delta_{0}(x)=0$ in Eq. (\ref{EqYRZ}), which has been observed by ARPES and interpreted as signature of particle-hole asymmetry \cite{Hashimoto2010, He2011, Yang2011}. As the momentum cut moves towards the nodal region, the lower energy band shifts up to close the spectral gap at the Fermi energy, which is consistent with the ARPES measurement \cite{He2011} and, as shown by Lee \cite{Lee2014}, cannot be fully explained in a CDW-induced-pseudogap scenario.

\section{Fermi surface reconstruction due to incommensurate CDW order} \label{SecFS}

\begin{figure}
\centering
\includegraphics[width=0.48\textwidth]{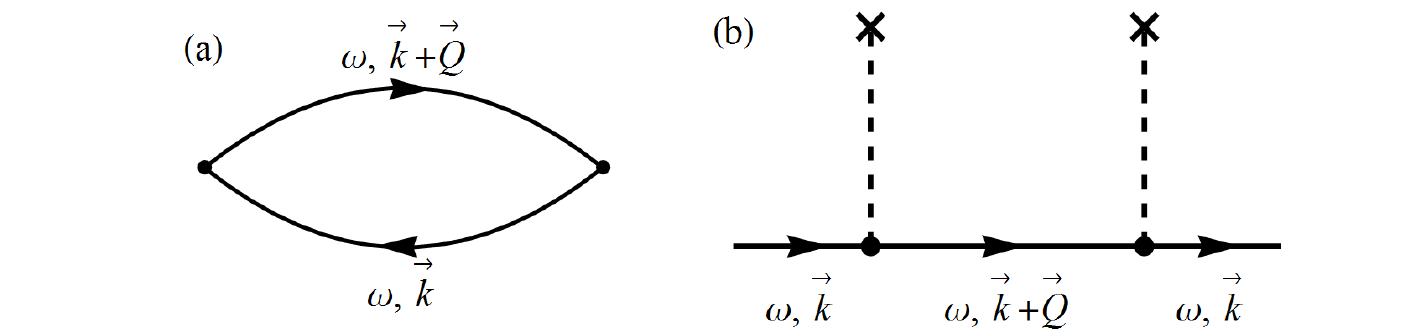}
\caption{(Color online) (a) The particle-hole bubble diagram for computing the CDW susceptibility $\chi_{\mathrm{CDW}}(\vec{Q})$. For the $s$- and $d$-form CDW, the vertices are multiplied by $1$ and $\cos (k_{x}+Q_{x}/2)-\cos (k_{y}+Q_{y}/2)$ respectively. (b) The self-energy correction from the CDW order perturbation up to $P_{0}^{2}$ order. For the $d$-form CDW order, each vertex contributes a factor $P_{0}(\cos (k_{x}+Q_{x}/2)-\cos (k_{y}+Q_{y}/2))$.}
\label{FigFeynman}
\end{figure}

It is well-established that the underdoped non-La-based cuprates exhibit bidirectional incommensurate CDW order at wavevectors $\vec{Q}_{1}=(Q, 0)$ and $\vec{Q}_{2}=(0,Q)$ with $Q/2\pi \simeq 0.3$. Recently, it is shown by Comin \emph{et al} \cite{Comin2014b} that $\vec{Q}_{i}$'s are the wavevectors connecting the tips of the Fermi arcs (the hotspots) by combining the ARPES and the resonant X-ray scattering (REXS) measurements. They also extracted the CDW wavevectors from the local maxima of the static CDW susceptibility of the YRZ Green's function along the momentum cuts $(Q_{x},0)$ and $(0,Q_{y})$ and found quantitative agreement with experiments. Therefore, we take this approach to locate the CDW wavevectors connecting the hotspots and study the induced Fermi surface reconstruction and quantum oscillation properties.

\begin{figure}
\centering
\includegraphics[width=0.48\textwidth]{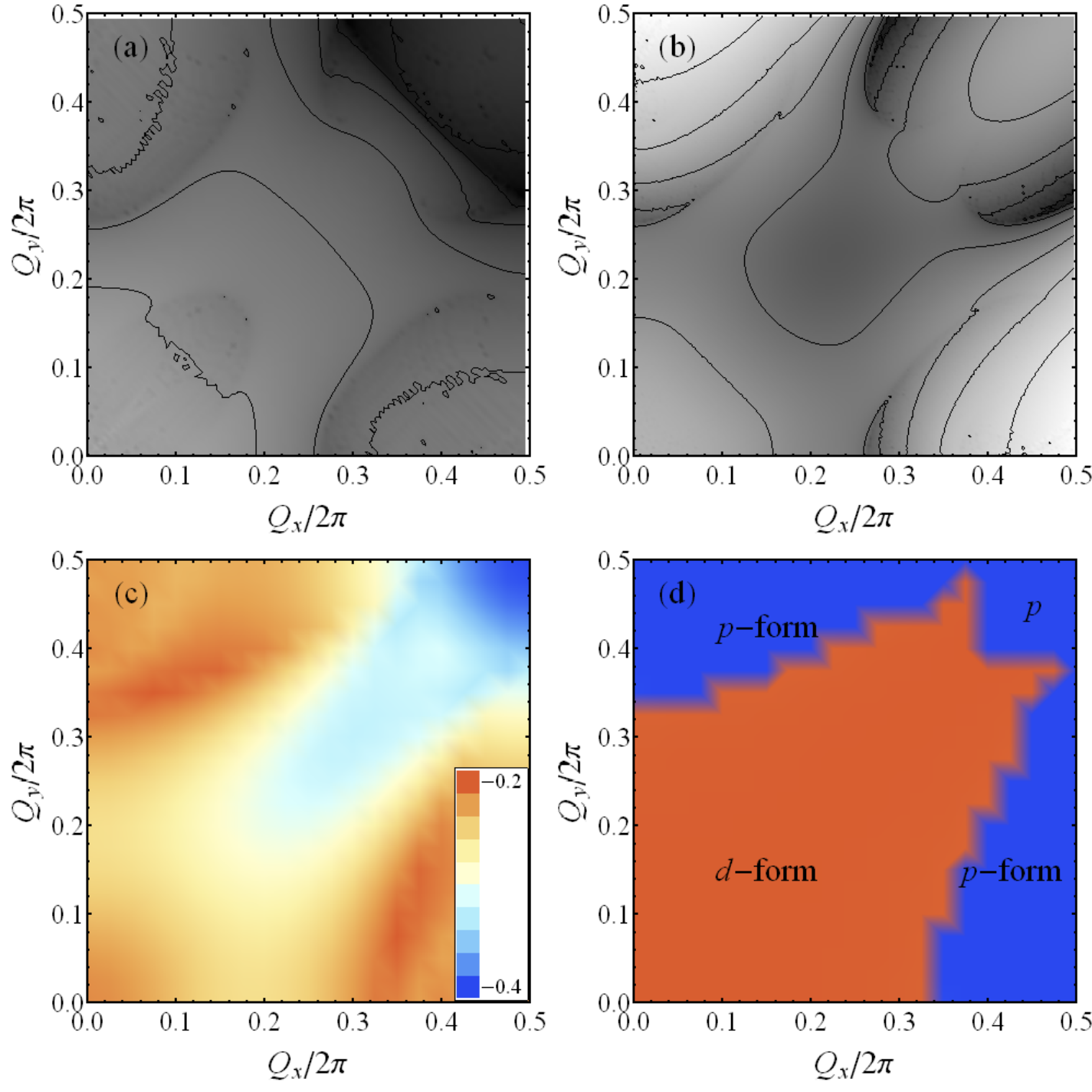}
\caption{(Color online) Contour plot of the (a) $s$- and (b) $d$-form static CDW susceptibilities $\chi_{\mathrm{CDW}}(\vec{Q})$ of the YRZ Green's function with doping $x=0.12$. (c) The lowest eigenvalues of the Hartree-Fock self-energy kernel in Eq. \ref{EqFree} indicating the CDW instability at each wavevector. (d) The overlap between the normalized CDW form factor $\Delta_{\vec{Q}}(\vec{k})$ and the standard $d$- and $p$- form basis functions in the momentum space. The region in orange is dominated by $d_{x^{2}-y^{2}}$-form and the regions in blue dominated by $p_{x}$- or $p_{y}$-form.}
\label{FigSusc}
\end{figure}

We calculate the static CDW susceptibility at zero temperature $\chi_{\mathrm{CDW}}(\vec{Q})$ of the YRZ Green's function by the particle-hole bubble diagram in Fig. \ref{FigFeynman} (a),
\begin{equation}
\chi_{\mathrm{CDW}}(\vec{Q})= -\int d\omega d^{2}\vec{k} G_{0}(\omega, \vec{k})G_{0}(\omega, \vec{k}+\vec{Q}) F(\vec{k}, \vec{k}+\vec{Q})^{2},
\end{equation}
in which $F(\vec{k},\vec{k}+\vec{Q}) = 1$ for $s$-form CDW order, i.e., local charge density modulation, and $F(\vec{k}, \vec{k}+\vec{Q}) = \cos (k_{x}+Q_{x}/2)\pm \cos (k_{y}+Q_{y}/2)$ for extended $s$-form and $d$-form CDW order, i.e., nearest-neighbor-bond-centered charge modulation \cite{Sachdev2013a}. Similar calculations for the $s$-form CDW susceptibility at finite temperature were also carried out in Refs. \onlinecite{Comin2014b} and \onlinecite{Scherpelz2014}, which yielded siminar results. The $s$- and $d$-form susceptibilities at $x=0.12$ are shown in Figs. \ref{FigSusc} (a) and (b). Two local maxima appear at $(Q,0)$ and $(0,Q)$, with $Q/2\pi\simeq 0.275$, corresponding to the wavevectors connecting the hotspots as indicated by the arrows in Fig. \ref{FigYRZSpectral} (a).

We also carry out the unconstrained Hartree-Fock calculations for the YRZ Green's function to find the CDW order instability in the presence of the short-range antiferromagnetic (AF) exhange interaction. The formalism was developed in Ref. \onlinecite{Sachdev2013a} and we follow its notations below. Suppose that the system develops an CDW order described by the following perturbation term in the mean field Hamiltonian,
\begin{equation}
H'=-\sum_{i,j}\Delta_{ij}c_{i\sigma}^{\dag}c_{j\sigma},
\end{equation}
The nonlocal charge order parameter $\Delta_{ij}$ can be Fourier transformed into the momentum space,
\begin{equation}
\Delta_{ij}=\frac{1}{V}\sum_{\vec{Q}}\sum_{\vec{k}}e^{i\vec{k}\cdot (\vec{r}_{i}-\vec{r}_{j})}\Delta_{\vec{Q}}(\vec{k})e^{i\vec{Q}\cdot (\vec{r}_{i}+\vec{r}_{j})/2},
\end{equation}
in which $V$ is the system volume. The form of the CDW order $\Delta_{\vec{Q}}(\vec{k})$ is not assumed in advance; instead, it is determined by lowering the free energy of the system as follows. The free energy $\Delta F$ in the presence of the CDW order expanded to the second order of $\Delta_{\vec{Q}}(\vec{k})$ is given by \cite{Sachdev2013a},
\begin{equation}
\Delta F=\sum_{\vec{k},\vec{k}',\vec{Q}'}\Delta_{\vec{Q}}^{*}(\vec{k})\sqrt{\Pi_{\vec{Q}}(\vec{k})}\mathcal{M}_{\vec{Q}}(\vec{k},\vec{k}')\sqrt{\Pi_{\vec{Q}}(\vec{k}')}\Delta_{\vec{Q}}(\vec{k}'),
\label{EqFree}
\end{equation}
in which the kernel $\mathcal{M}_{\vec{Q}}(\vec{k},\vec{k}')$ is given by
\begin{equation}
\mathcal{M}_{\vec{Q}}(\vec{k},\vec{k}')=\delta_{\vec{k},\vec{k}'}+\frac{3}{V}\chi_{0}(\vec{k}-\vec{k}')\sqrt{\Pi_{\vec{Q}}(\vec{k})\Pi_{\vec{Q}}(\vec{k}')}.
\end{equation}
The polarizability $\Pi_{\vec{Q}}(\vec{k})$ is given by
\begin{equation}
\Pi_{\vec{Q}}(\vec{k})=-\sum_{i\omega_{n}}G_{0}(i\omega_{n},\vec{k}+\vec{Q}/2)G_{0}(i\omega_{n},\vec{k}-\vec{Q}/2), \label{EqPolar}
\end{equation}
in which the summation is taken over the Matsubara frequency $\omega_{n}=2\pi n k_{\mathrm{B}}T$ at finite temperature $T$. The interaction vertex factor $\chi_{0}(\vec{q})$ for the nearest neighbor AF exchange coupling $J\sum_{\langle ij\rangle} \vec{S}\cdot \vec{S}_{j}$ is given by
\begin{equation}
\chi_{0}(\vec{q})=-\frac{1}{2}J(\cos q_{x}+\cos q_{y}).
\end{equation}

Given the free energy expression, Eq. (\ref{EqFree}), the strongest CDW instability at each wavevector $\vec{Q}$ sets in for the form factor $\Delta_{\vec{Q}}(\vec{k})$ being proportional to $\phi_{\vec{Q}}(\vec{k})/\sqrt{\Pi_{\vec{Q}}(\vec{k})}$, in which $\phi_{\vec{Q}}(\vec{k})$ is the eigenvector of the kernel $\mathcal{M}_{\vec{Q}}(\vec{k},\vec{k}')$ with the lowest eigenvalue, so we solve the eigenvalue equation
\begin{equation}
\frac{1}{V}\sum_{\vec{k}'}\sqrt{\Pi_{\vec{Q}}(\vec{k})}\chi_{0}(\vec{k}-\vec{k}')\sqrt{\Pi_{\vec{Q}}(\vec{k}')}\phi_{\vec{Q}}(\vec{k}')=\lambda_{\vec{Q}}\phi_{\vec{Q}}(\vec{k}),
\end{equation}
on a lattice with $80\times 80$ sites at $k_{\mathrm{B}}T=0.01t$ for the lowest eigenvalue $\lambda_{\vec{Q}}$ and decompose $\Delta_{\vec{Q}}(\vec{k})\propto \phi_{\vec{Q}}(\vec{k})/\sqrt{\Pi_{\vec{Q}}(\vec{k})}$ into the superposition of the standard orthonormal basis functions \cite{Sachdev2013a}, e.g., $1$ ($s$-form), $\cos k_{x}\pm \cos k_{y}$ (extended $s$- and $d_{x^{2}-y^{2}}$-form) and $\sqrt{2}\sin k_{x,y}$ ($p_{x,y}$-form).

\begin{figure}
\centering
\includegraphics[width=0.48\textwidth]{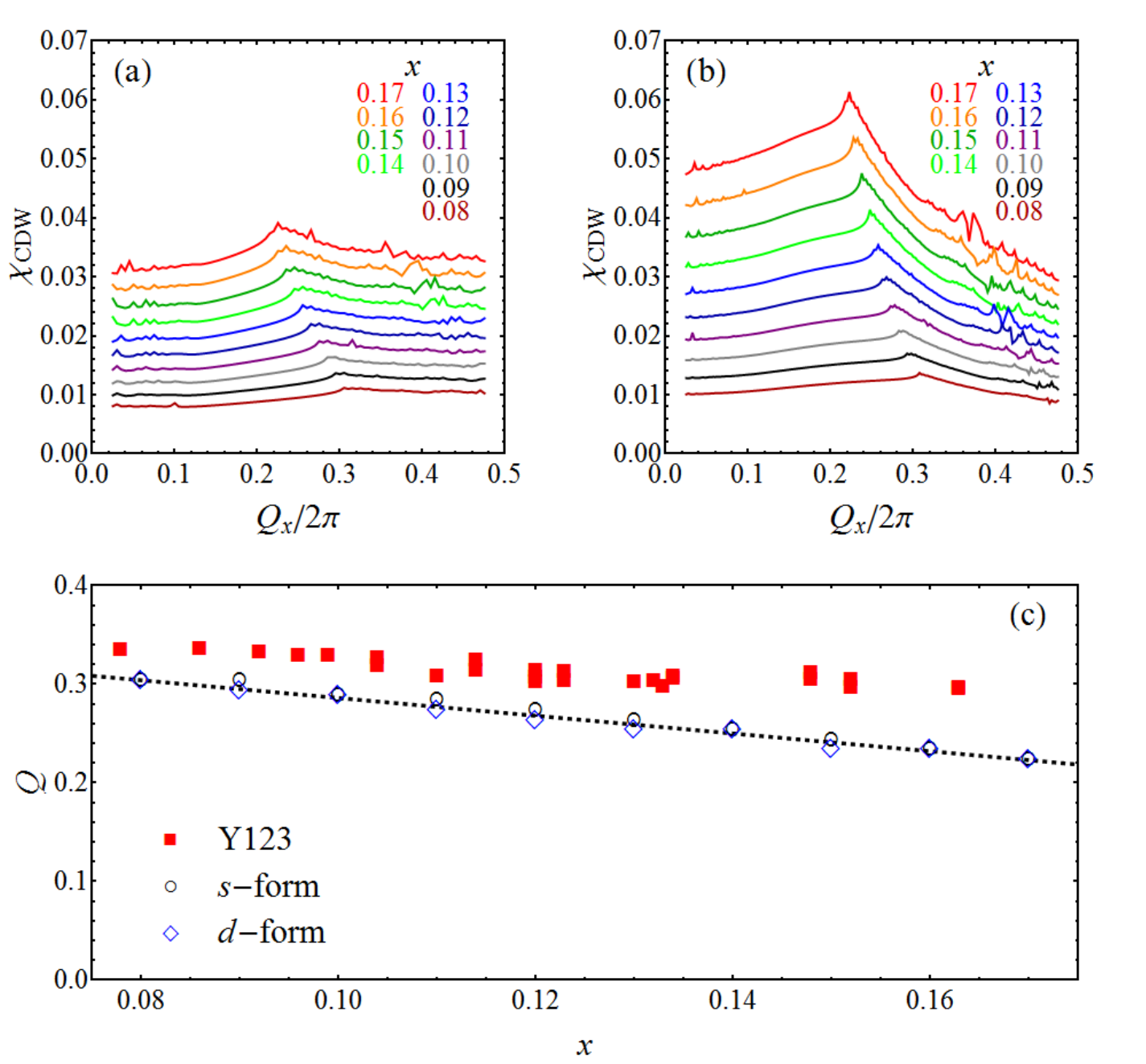}
\caption{(Color online) The doping dependence of (a) $s$- and (b) $d$-form CDW susceptibilities $\chi_{\mathrm{CDW}}(\vec{Q})$ along the momentum cut $Q_{y}=0$. (c) The doping dependence of the peak positions for $s$-form (black circles) and $d$-form (blue diamonds) CDW susceptibilities. The experiment results taken from Refs. \onlinecite{Tabis2014, Blanco-Canosa2014, Huecker2014} are included for comparison. The dashed line is guide to the eyes.}
\label{FigQvsx}
\end{figure}

The lowest eigenvalue $\lambda_{\vec{Q}}$ indicating the CDW instability at each wavevector $\vec{Q}$ is shown in Fig. \ref{FigSusc} (c). Although the global minimum appears around $(\pi, \pi)$, two local minima also show up at $(Q, 0)$ and $(0, Q)$, corresponding to the enhanced scattering between the hotspots. As shown in Fig. \ref{FigSusc} (d), the CDW instability at these wavevectors are dominated by $d$-form, which is consistent with experiments \cite{Fujita2014, Comin2014}. The doping dependence of $\vec{Q}$ is shown in Fig. \ref{FigQvsx} and in good agreement with the experiments in YBa$_{2}$Cu$_{3}$O$_{6+\delta}$. Therefore, we focus on the incommensurate $d$-form CDW order at the wavevectors connecting the hotspots and study the induced Fermi surface reconstruction and the quantum oscillation in the rest of this work.

We make a few comments on the Hartree-Fock calculations. In our results shown in Fig. \ref{FigSusc} (c), the CDW instability near $(\pi, \pi)$ is stronger than that connecting the hotspot for the YRZ state, due to the nearly nested (parallel) Fermi surface patches by shifting by $(\pi, \pi)$, because the self-energy diverges precisely at the \emph{commensurate} magnetic Brillouin zone boundary $k_{x}\pm k_{y}=\pm \pi$ in the phenomenological YRZ ansatz. If such a constraint on the self-energy divergence line is relaxed, the YRZ ansatz is equivalent to the fractionalized Fermi liquid (FL$^{*}$) state proposed by Sachdev and collaborators \cite{Qi2010a, Moon2011}, which was derived for itinerant electrons coupled to short-range AF order. It is found in Ref. \onlinecite{Chowdhury2014} that the FL$^{*}$ state also exhibits two sets of local maxima of the CDW instability, and that at the wavevectors connecting the hotspots is stronger than that near $(\pi, \pi)$ for short-range AF coupling. Therefore, the relative strength of CDW instability is sensitive to the parameter choice of the electron structure and the interactions, which may account for the diversity of CDW forms in different cuprate families. In the rest of this work, we focus on the bidirectional CDW in the non-La-based compounds with the wavevectors connecting the hotspots. The scenario of Fermi surface reconstruction by incommensurate CDW order is expected to hold true for the FL$^{*}$ state \cite{Chowdhury2014} as well as other postulated pseudogap states with nodal hole-like Fermi pockets.

The Fermi surface is reconstructed when the \emph{static} CDW order sets in at low temperature. The CDW order is described by introducing the following perturbation term into the Green's function,
\begin{equation} \label{EqCDW}
H_{\mathrm{CDW}}= \sum_{\vec{k},\sigma}P(\vec{k})\sum_{i=1,2}c_{\vec{k}+\vec{Q}_{i}/2,\sigma}^{\dag}c_{\vec{k}-\vec{Q}_{i}/2,\sigma},
\end{equation}
in which the $d$-form CDW order parameter is given by $P(\vec{k})=P_{0}(\cos k_{x}-\cos k_{y})$. The wavevectors $\vec{Q}_{1}=(Q,0)$ and $\vec{Q}_{2}=(0,Q)$ are taken as those connecting the hotspots as discussed above. It leads to the self-energy correction up to $P_{0}^{2}$ order as shown in Fig. \ref{FigFeynman} (b),
\begin{equation}
\Sigma_{\mathrm{CDW}}(\omega, \vec{k}) = \sum_{i=1,2}P(\vec{k}+\vec{Q}_{i}/2)^{2}G_{0}(\omega, \vec{k}+\vec{Q}_{i}).
\end{equation}

The CDW-perturbed spectral function is shown in Figs. \ref{FigQOvsDoping} (a) and (b). A spectral gap opens around the hotspots, thus the Fermi surface is reconstructed. When subject to magnetic fields, an electron wave packet moves along the Fermi surface in the semiclassical theory \cite{Ziman1972principles}. In the presence of the CDW order, the electron can be scattered at the hotspot to another patch of Fermi surface, then it continues moving until getting scattered again at another hotspot. Therefore, the semiclassical trajectory forms a magnetic orbit composed of the Fermi surface patches joined up at the hotspots. In the first quadrant in Fig. \ref{FigQOvsDoping} (a), we shift the Fermi surface patches by $\vec{Q}_{i}$ and join them up to illustrate the semiclassical closed magnetic orbits of the electrons in the presence of the CDW order. The inner patches (the Fermi arcs) form an electron-like Fermi pocket, denoted as the $\alpha$ orbit, which accounts for the negative Hall and Seebeck coefficients in experiments \cite{LeBoeuf2007, Chang2010, Doiron-Leyraud2013}. The outer patches with vanishingly small spectral weight (the ``shadow'' patches) also join up to form a new hole-like Fermi pocket, denoted as the $\beta$ orbit.

The area $S$ of the Fermi pocket enclosed by the $\alpha$ orbit is calculated by numerical integration. It changes systematically with the doping concentration as given in Table \ref{TabTheo} and illustrated in Fig. \ref{FigQOvsDoping} (c). The results extracted from the quantum oscillation experiments according to the Onsager relation \cite{Ziman1972principles},
\begin{equation}
F=\frac{\hbar c}{4\pi^2 e}S,
\end{equation}
are also included for comparison. The $\alpha$ orbit areas from our calculations are in quantitative agreement with the dominant  oscillation frequency in experiments. The $\beta$ orbit can explain the high-frequency peak observed by Sebastian \textit{et al} with an about three times higher frequency than the dominant peak \cite{Sebastian2008b, Sebastian2010, Sebastian2010b, Sebastian2011}. This large-frequency peak is not observed by other groups \cite{Audouard2009}, which may be due to the vanishingly small spectral weight on this pocket. We note that this peak was also attributed to the ortho-II potential in the YBa$_{2}$Cu$_{3}$O$_{6+\delta}$ materials \cite{Podolsky2008}.

\section{Density of states oscillation in magnetic field} \label{SecQO}

\begin{table}
\centering
\caption{The doping dependence of the electron pocket area derived from the semiclassical analysis and the DoS oscillation.}
\label{TabTheo}
\begin{tabular}{c|c|c}
\hline \hline
Doping $x$ & Semiclassical $S/S_{\mathrm{BZ}}$ & DoS oscillation $S/S_{\mathrm{BZ}}$ \\ 
\hline \hline
0.08 & 0.0152 & 0.0152 \\ 
0.09 & 0.0140 & 0.0137 \\ 
0.10 & 0.0191 & 0.0189 \\ 
0.11 & 0.0200 & 0.0201 \\ 
0.12 & 0.0237 & 0.0238 \\ 
0.13 & 0.0273 & 0.0274 \\ 
0.14 & 0.0287 & 0.0323 \\ 
0.15 & 0.0363 & 0.0366 \\ 
\hline \hline
\end{tabular} 
\end{table}

\begin{figure}
\centering
\includegraphics[width=0.48\textwidth]{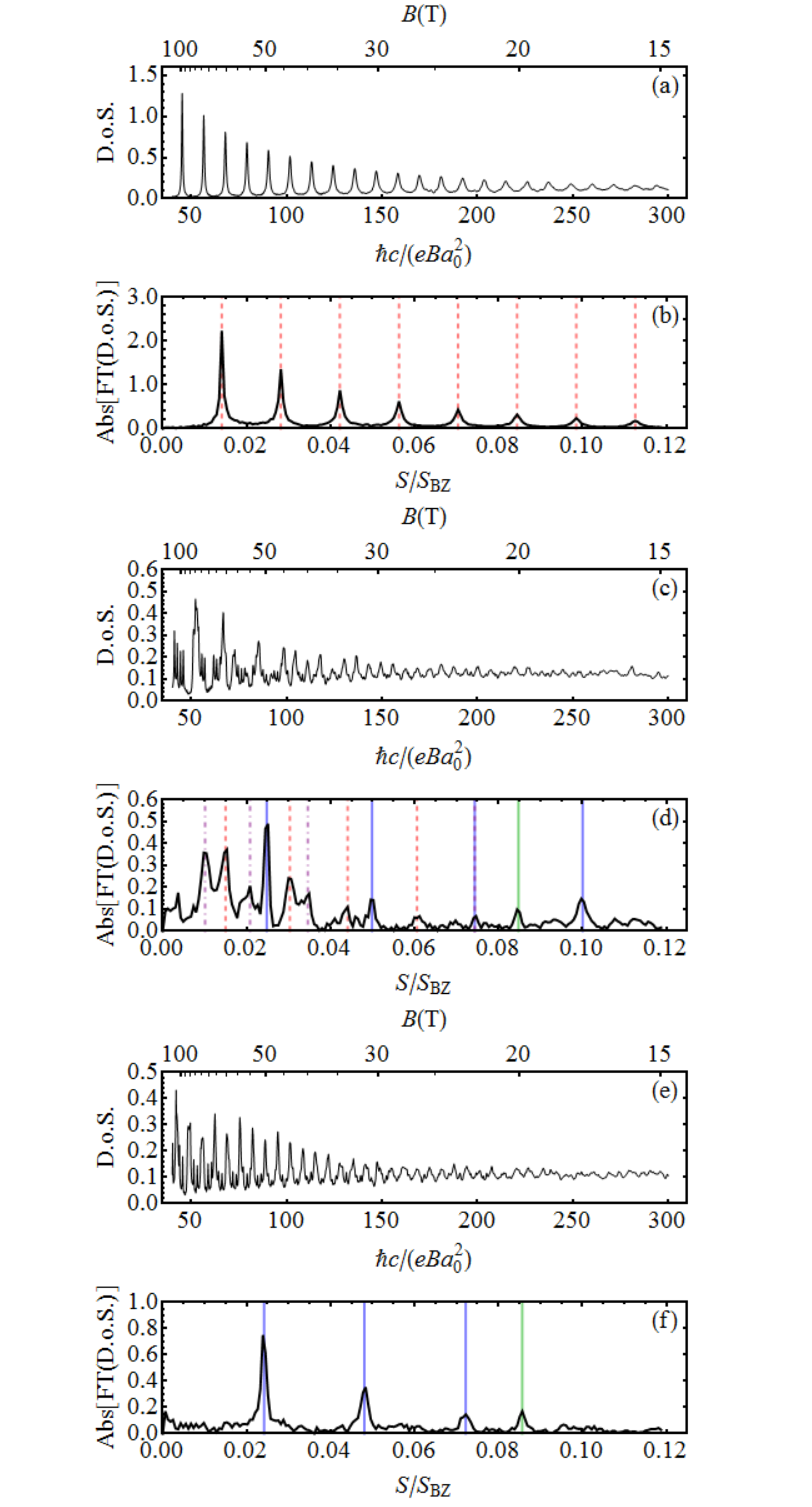}
\caption{(Color online) The calculated density of states at the Fermi energy in magnetic field and its Fourier transform for the CDW order magnitude (a, b) $P_{0}=0$, (c, d) $0.2$ and (e, f) $0.3$ on a lattice of $400\times 80$ sites. The horizontal axis of the Fourier transform has been converted into the Fermi pocket area using the Onsager relation. The dashed red lines indicate the oscillation frequency corresponding to the original hole pockets in YRZ Green's function and its multiples, the solid blue lines for the reconstructed electron-like $\alpha$ orbit and the green lines for the new hole-like $\beta$ orbit. The dotdashed purple lines indicate the higher harmonics from both $\alpha$ and $\beta$ orbits.}
\label{FigDoSOscillation}
\end{figure}

In order to corroborate the semiclassical analysis and to clarify the impact of the vanishingly small spectral weight of the $\beta$ orbit on the quantum oscillation, we directly calculate the density of states (DoS) at the Fermi energy in magnetic field by introducing the following effective Hamiltonian
\begin{equation} \label{EqEff}
H_{0}^{\mathrm{eff}}=\sum_{\vec{k},\sigma}
\begin{pmatrix}
c_{\vec{k}\sigma}^{\dag}&\tilde{c}_{\vec{k}\sigma}^{\dag}
\end{pmatrix}
\begin{pmatrix}
\xi(\vec{k})& \Delta(\vec{k})\\
\Delta(\vec{k})& -\xi_{0}(\vec{k})
\end{pmatrix}
\begin{pmatrix}
c_{\vec{k}\sigma}\\
\tilde{c}_{\vec{k}\sigma}
\end{pmatrix},
\end{equation}
in which $c_{\vec{k}\sigma}$ and $\tilde{c}_{\vec{k}\sigma}$ denote the annihilation operators of the physical electron band and an auxiliary band respectively. By projecting onto the physical electron ($c$-electron) band, the effective Hamiltonian reproduces the YRZ Green's function up to a constant factor,
\begin{equation}
\begin{split}
G_{0}^{\mathrm{eff}}(\omega, \vec{k})&=-i\int dt e^{-i\omega t}\langle \mathcal{T}_{t} c_{\vec{k}\sigma}(t)c^{\dag}_{\vec{k}\sigma}(0)\rangle\\
&=g_{t}(x)^{-1}G_{0}(\omega, \vec{k}).
\end{split}
\end{equation}

The effective Hamiltonian enables us to calculate the electron DoS on a lattice in the presence of the CDW order and magnetic field, $D(\epsilon_{\mathrm{F}})=\frac{1}{\pi}\mathrm{Im}\mathrm{Tr}\hat{P}\frac{1}{H-i\eta}\hat{P}$, in which $H=H_{0}^{\mathrm{eff}}+H_{\mathrm{CDW}}^{\mathrm{eff}}$ is the effective Hamiltonian of the CDW-perturbed YRZ state on a lattice given below and $\hat{P}$ denotes the projection operator onto the $c$-electron band.

In the real space, the effective Hamiltonian in magnetic field reads
\begin{equation} \label{EqHeffReal}
\begin{split}
H_{0}^{\mathrm{eff}} =&  -\sum_{i,j,\sigma}t_{ij}(x)c_{i\sigma}^{\dag}c_{j\sigma}e^{-i e A_{ij}^{e}}+t(x)\sum_{\langle ij\rangle,\sigma}\tilde{c}_{i\sigma}^{\dag}\tilde{c}_{j\sigma}e^{-i e A_{ij}^{e}}\\
&+\Delta_{0}(x)\sum_{i,\sigma}(c_{i\sigma}^{\dag}\tilde{c}_{i+\hat{x},\sigma}e^{-i e A_{i,i+\hat{x}}^{e}}-c_{i\sigma}^{\dag}\tilde{c}_{i+\hat{y},\sigma}e^{-i e A_{i,i+\hat{y}}^{e}})\\
&+\mathrm{H.c.}-\mu(x)\sum_{i,\sigma}c_{i\sigma}^{\dag}c_{i\sigma},
\end{split}
\end{equation}
in which $t_{ij}(x)=t(x)$, $t'(x)$ and $t''(x)$ for $i$ and $j$ being the first, second and third nearest neighbors, respectively. We choose the Landau gauge for the electromagnetic vector potential in our calculations, $A^{e}_{i,i+\hat{y}}=\phi x_{i}$, $A^{e}_{i,i+\hat{x}}=0$, in which $\phi=Ba_{0}^{2}$ is the magnetic flux through each plaquette.

The CDW order of the $c$-electrons in the effective Hamiltonian approach is given by
\begin{equation} \label{EqHCDW}
\begin{split}
H_{\mathrm{CDW}}^{\mathrm{eff}}=& g_{t}(x)P_{0}\sum_{i,\sigma}\Big( \cos[Q(x_{i}+1/2)]c_{i\sigma}^{\dag}c_{i+\hat{x},\sigma}\\
& -\cos(Qx_{i}) c_{i\sigma}^{\dag}c_{i+\hat{y},\sigma}e^{-ieA^{e}_{i,i+\hat{y}}} + \cos(Q y_{i})c_{i\sigma}^{\dag}c_{i+\hat{x},\sigma}\\
&-\cos[Q(y_{i}+1/2)]c_{i\sigma}^{\dag}c_{i+\hat{y},\sigma}e^{-ieA^{e}_{i,i+\hat{y}}}\Big)+\mathrm{H.c.},
\end{split}
\end{equation}
in which an extra factor $g_{t}(x)$ is included in front of the CDW order magnitude $P_{0}$ so that the CDW-perturbed Green's functions derived from the YRZ Green's function [Fig. \ref{FigFeynman} (b)] and the effective Hamiltonian are identical up to a constant.

The real-space Hamiltonian is put on a lattice with $N_{x}\times N_{y}$ sites. We adopt periodic boundary condition along the $y$ direction with $N_{y}=80$, which poses a mild commensurate constraint on the CDW wavevectors, $N_{y}Q/2\pi=\mathrm{integer}$. We adopt open boundary condition along the $x$ direction, so the total magnetic flux is not quantized on the cylinder and the magnetic field can be tuned continuously. We choose $N_{x}=400$, which is large enough so that the finite size effect is negligible. As a consistency check, calculations on a lattice with $200\times 200$ sites are also performed and nearly identical results are found. The lattice site indices are ordered such that the Hamiltonian is block-tridiagonal with $N_{x}/2\times N_{x}/2$ blocks and each block $4N_{y}$-dimensional. The efficient iterative algorithm introduced by Allais \textit{et al} \cite{Allais2014} (see Appendix \ref{AppAlgorithm}) is adopted to calculate the diagonal blocks of $(H-i\eta)^{-1}$ ($\eta=0.001t$ is a Lorentzian broadening) and the $c$-electron DoS is calculated by taking the trace of $(H-i\eta)^{-1}$ only over the $c$-electron sector.

The results for doping $x=0.12$ are shown in Fig. \ref{FigDoSOscillation}. In the absence of the CDW order, $P_{0}=0$, the original hole pockets in the YRZ Green's function yield the DoS oscillation as shown in Figs. \ref{FigDoSOscillation} (a) and (b). The Fermi pocket area $S/S_{\mathrm{BZ}}=0.0141$ ($S_{\mathrm{BZ}}=4\pi^2/a_{0}^{2}$) derived from the Onsager relation agrees with the semiclassical analysis $S/S_{\mathrm{BZ}}=x/8=0.015$. As we turn on the CDW order, the DoS oscillation frequency spectrum shows new peaks corresponding to the reconstructed Fermi pockets, and the original peaks gradually diminish, as shown in Figs. \ref{FigDoSOscillation} (c)--(f) for $P_{0}=0.2$ and $0.3$ respectively. In Fig. \ref{FigDoSOscillation} (f), the DoS oscillation is dominated by the reconstructed $\alpha$ orbit and its peak corresponds to a pocket area $S/S_{\mathrm{BZ}}=0.0238$, which perfectly matches the electron pocket area $0.0237$ in the semiclassical analysis. The DoS oscillations are also calculated for other doping concentrations and the extracted $\alpha$ orbit areas are listed in Table \ref{TabTheo}. They are in good agreement with the semiclassical results as well as the experiments, as shown in Fig. \ref{FigQOvsDoping} (c).

Except for the dominant peak and its multiples, we also find a high-frequency peak, which corresponds to the new hole-like $\beta$ orbit discussed in Sec. \ref{SecFS}. Its area, $S/S_{\mathrm{BZ}}=0.0857$, equals the $\alpha$ orbit area plus those of the four YRZ hole pockets ($x/2$ in total). However, this peak is much lower than that of the $\alpha$ orbit due to the vanishingly small spectral weight near the nodal points. This may explain the controversy in experiments \cite{Sebastian2008b, Sebastian2010, Sebastian2011, Sebastian2012a, Audouard2009}. More experiments are needed to confirm this large-frequency $\beta$ orbit.

\subsection{Robustness against local disorder}

In order to check the robustness of the magnetic orbits against local disorder, which is unavoidable in real materials, we impose $5\%$ randomness to the chemical potential in the effective Hamiltonian, i.e., replacing $\mu(x)$ in Eq. (\ref{EqHeffReal}) with $\mu(x)(1+\delta_{i})$ at each site, in which $\delta_{i}$ is uniformly distributed in $[-0.05,0.05]$. We find that the DoS oscillation spectrum shown in Figs. \ref{FigDisorder} (a) and (b) has little change as compared with Figs. 5 (e) and (f) (without disorder). This demonstrates that the quantum oscillations from both $\alpha$ and $\beta$ orbits are robust against weak local disorder.

\subsection{Robustness against static CDW fluctuations}

\begin{figure}
\centering
\includegraphics[width=0.48\textwidth]{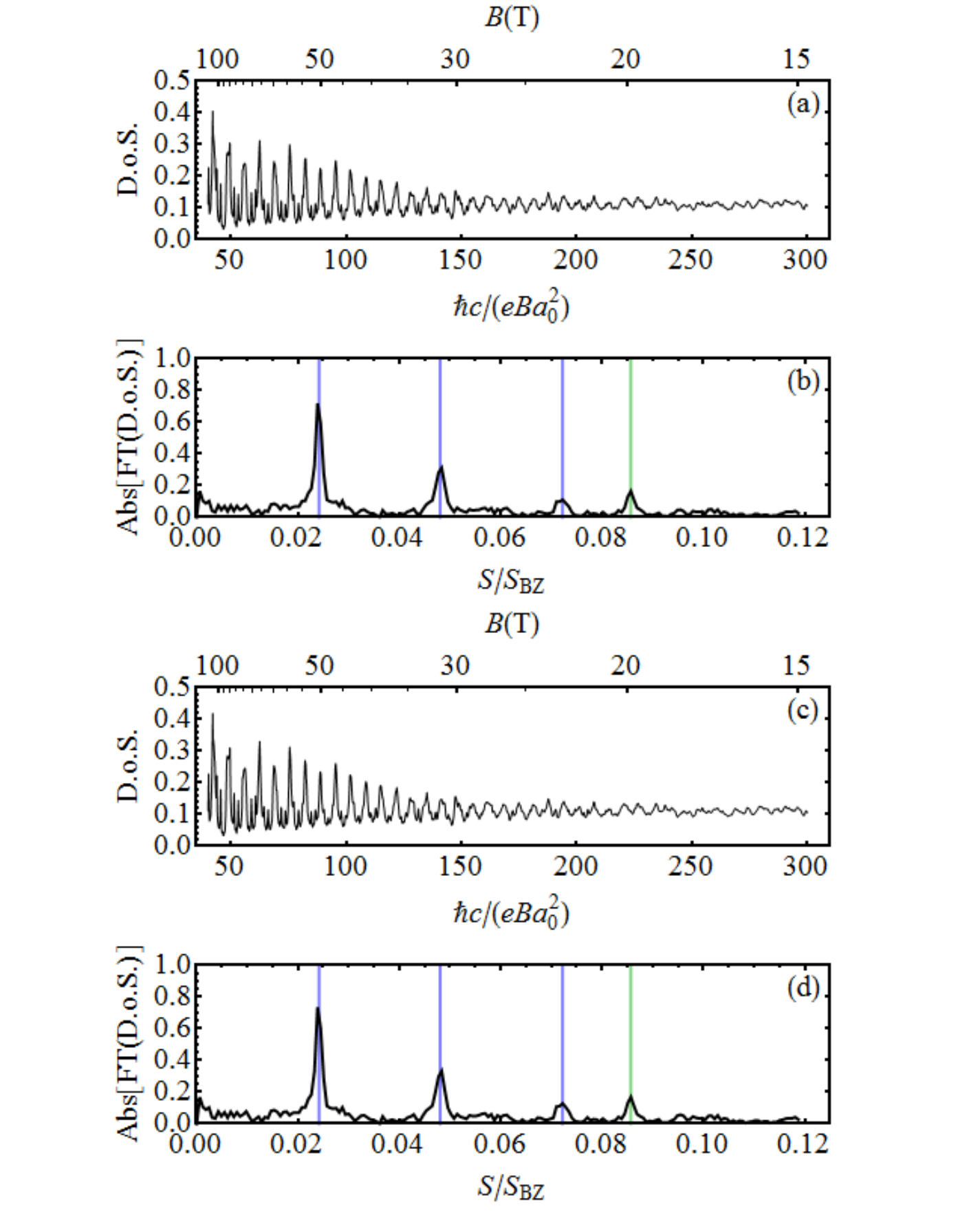}
\caption{(Color online) The density of states at the Fermi energy in magnetic field and the Fourier transform for CDW order magnitude $P_{0}=0.3$ in the presence of (a, b) $5\%$ randomness imposed on the $c$-electron chemical potential $\mu(x)$ and (c, d) $10\%$ randomness imposed on the bond-centered CDW order magnitude $P_{0}$ on a $400\times 80$ lattice. The horizontal axis of the Fourier transform has been converted into the Fermi pocket area using the Onsager relation. The solid blue line indicates the oscillation frequency corresponding to the reconstructed electron pocket while the green line for the new hole pocket.}
\label{FigDisorder}
\end{figure}

We also consider the robustness of the quantum oscillation against CDW fluctuations. Because our calculations of the DoS in magnetic field rely on the quadratic form of the Hamiltonian, Eqs. (\ref{EqHeffReal}) and (\ref{EqHCDW}), it is inaccessible for us to study the generic dynamical CDW fluctuations, which are usually controlled by electron interaction terms. Instead, we introduce quenched randomness to the bond-centered CDW order parameter, i.e., by replacing $P_{0}$ in Eq. (\ref{EqHCDW}) with $P_{0}(1+\delta_{ij})$ on each nearest neighbor bond, to check the robustness of DoS oscillation against the \emph{static} CDW fluctuations. $\delta_{ij}$ is uniformly distributed in $[-0.1,0.1]$. The results are shown in Figs. \ref{FigDisorder} (c) and (d). The quantum oscillations from both $\alpha$ and $\beta$ orbits are robust against the static CDW fluctuations.

\section{Summary} \label{SecSum}

In this work, we have considered the phenomenological synthesis of the doped RVB state and the incommensurate CDW order for the underdoped cuprates. Starting from the YRZ ansatz of the single-particle Green's function and introducing the incommensurate CDW order at the wavevectors connecting the tips of the Fermi arcs (the hotspots), we find that the Fermi arcs join up to form an electron-like Fermi pocket, which is confirmed by the DoS oscillation in magnetic field. The doping dependence of the electron pocket area is in quantitative agreement with experiments. We also find a new hole-like Fermi pocket formed by joining the outer ``shadow'' patches of the original hole pockets, with its area equal to that of the electron pocket plus those of the original hole pockets. This new hole pocket can explain the high-frequency peak observed by Sebastian \textit{et al} \cite{Sebastian2008b, Sebastian2010, Sebastian2010b, Sebastian2011}. The controversy in experiments \cite{Audouard2009} may be due to the small magnitude of this oscillation peak because of the vanishingly small spectral weight on the outer patches. Further confirmation of this large-frequency oscillation peak can be taken as evidence of the ``shadow'' side of the nodal hole pockets in the pseudogap regime.

This formalism provides several tunable parameters, e.g., the CDW order magnitude $P_{0}$ and the Lorentzian broadening $\eta$, which acts as an efficient temperature in the DoS calculations, so we may gain more insight into the quantum oscillations in underdoped cuprates. In particular, for the moderate $P_{0}$, the DoS in magnetic field exhibit a rich multi-component oscillation pattern, as shown in Figs. \ref{FigDoSOscillation} (c) and (d). A detailed analysis is presented in a separate work \cite{Zhang2014c}.

\begin{acknowledgements}
We are grateful to helpful discussions with R.-H. He, S. Sachdev, Z.-Y. Weng, H. Yao and P. Ye.  T. M. Rice is especially acknowledged for enlightening suggestions.  L.Z. is supported by the National Basic Research Program of China (973 Program, No. 2010CB923003). Research at Perimeter Institute is supported by the Government of Canada through Industry Canada and by the Province of Ontario through the Ministry of Research (J.W.M.).
\end{acknowledgements}

\appendix

\section{Renormalized mean field theory} \label{AppRMFT}

The renormalized mean field theory (RMFT) was devised to study the $t$-$J$ model analytically by adopting the renormalization factors $g_{t,J}(x)$ from the Gutzwiller approximation to account for the single-occupancy condition \cite{Zhang1988}. The effective Hamiltonian in the unprojected Hilbert space with the renormalized factors are
\begin{equation}
H=-g_{t}t\sum\limits_{\langle ij\rangle}(c_{i\sigma}^{\dag} c_{j\sigma}+\mathrm{H.c.})+g_{J}J\sum\limits_{\langle ij\rangle} \vec{S}_{i}\cdot \vec{S}_{j},
\end{equation}
in which $g_{t}=2x/(1+x)$ and $g_{J}=4/(1+x)^{2}$ are the renormalization factors for the hopping and the AF exchange terms respectively.

\begin{figure}
\centering
\includegraphics[width=0.48\textwidth]{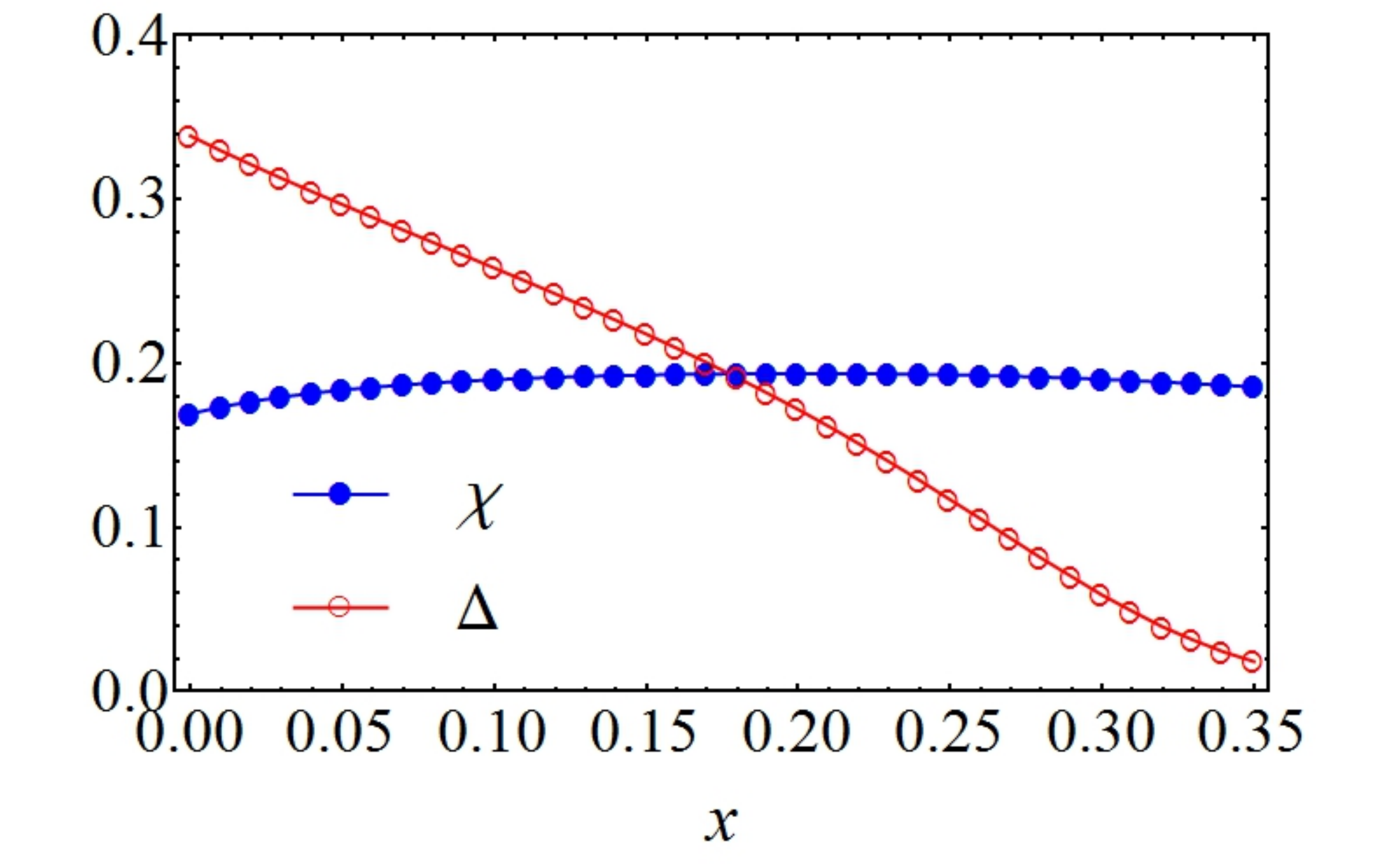}
\caption{(Color online) The hopping and paring parameters $\chi$ and $\Delta$ derived self-consistently from the renormalized mean field theory for $J=t/3$.}
\label{FigRMFT}
\end{figure}

Introducing the mean field parameters $\chi_{\hat{\tau}} = \langle c_{i\sigma}^{\dag} c_{i+\hat{\tau},\sigma}\rangle$ and $\Delta_{\hat{\tau}} = \sum_{\sigma}\langle\sigma c_{i\sigma}^{\dag}c_{i+\hat{\tau},-\sigma}^{\dag}\rangle$ to describe the electron hopping and RVB pairing amplitudes, in which $\hat{\tau}=\hat{x}, \hat{y}$, and assuming the $d$-wave RVB pairing, $\chi_{\hat{\tau}}=\chi$, $\Delta_{\hat{x}}=-\Delta_{\hat{y}}=\Delta$, we find the following mean field Hamiltonian
\begin{equation}
H_{\mathrm{MF}}=\sum_{\vec{k}}
\begin{pmatrix}
c_{\vec{k}\uparrow}^{\dag}& c_{-\vec{k}\downarrow}
\end{pmatrix}
\begin{pmatrix}
\xi_{0}(\vec{k})-\mu & -\Delta(\vec{k})\\
-\Delta(\vec{k}) & -\xi_{0}(\vec{k})+\mu
\end{pmatrix}
\begin{pmatrix}
c_{\vec{k}\uparrow}\\
c_{-\vec{k}\downarrow}^{\dag}
\end{pmatrix},
\end{equation}
in which $\xi_{0}(\vec{k})= -2t(x)(\cos k_{x}+\cos k_{y})$ and $\Delta(\vec{k})=\Delta_{0}(x)(\cos k_{x}-\cos k_{y})$ with $t(x)=g_{t}(x)t+3g_{J}(x)J\chi/8$ and $\Delta_{0}(x)=3 g_{J}(x)J\Delta/4$. The self-consistency equations are given by
\begin{align}
\chi &= \frac{1}{4N}\sum_{\vec{k}}\frac{1}{E_{\vec{k}}}(\mu-\xi_{0}(\vec{k}))(\cos k_{x}+\cos k_{y}),\\
\Delta &= \frac{3}{4N}\sum_{\vec{k}}\frac{1}{E_{\vec{k}}}g_{J}J\Delta (\cos k_{x}-\cos k_{y})^{2},\\
x &= \frac{1}{N}\sum_{\vec{k}}\frac{1}{E_{\vec{k}}}(\mu-\xi_{0}(\vec{k})),
\end{align}
in which $N$ is the lattice size and
\begin{equation}
E_{\vec{k}}=\sqrt{(\xi_{0}(\vec{k})-\mu)^{2}+\Delta(\vec{k})^{2}}
\end{equation}
is the mean field energy dispersion. These equations are solved for $J=t/3$ and the results are shown in Fig. \ref{FigRMFT}.

\section{Iterative algorithm in DoS calculations} \label{AppAlgorithm}

In order to calculate the density of states at the Fermi energy, $D(\epsilon_{\mathrm{F}})=\frac{1}{\pi}\mathrm{Im}\mathrm{Tr}\hat{P}\frac{1}{H-i\eta}\hat{P}$, we adopt the algorithm introduced by Allais \emph{et al} \cite{Allais2014} to calculate the diagonal blocks of $(H-i\eta)^{-1}$. On a lattice of $N_{x}\times N_{y}$ sites with open boundary condition along the $x$ direction, the two-band model up to the third-nearest-neighbor hopping terms, Eqs. (\ref{EqHeffReal}) and (\ref{EqHCDW}), can be arranged into the following block-tridiagonal form, with $N_{x}/2\times N_{x}/2$ blocks and each block $4N_{y}\times 4N_{y}$ dimensional (we follow the notations in Ref. \onlinecite{Allais2014}),
\begin{equation}
H-i\eta =
\begin{pmatrix}
h_{11}& t_{12}& 0& \ldots \\
t_{21}& h_{22}& t_{23}& \ldots \\
0& t_{32}& h_{33}& \ldots \\
\vdots& \vdots& \vdots& \ddots
\end{pmatrix}.
\end{equation}

The diagonal blocks $G_{ii}$ of $G=(H-i\eta)^{-1}$ can be calculated with the following iterative algorithm,
\begin{align*}
&L_{1}=0;\\
&\texttt{do}~i = 1: N_{x}/2-1\\
&~~~~~~~L_{i+1}=t_{i+1,i}(h_{ii}-L_{i})^{-1}t_{i,i+1};\\
&R_{N_{x}/2}=0;\\
&\texttt{do}~i = N_{x}/2: 2\\
&~~~~~~~R_{i-1}=t_{i-1,i}(h_{ii}-R_{i})^{-1}t_{i,i-1};\\
&\texttt{do}~i = 1: N_{x}/2\\
&~~~~~~~G_{ii}=(h_{ii}-L_{i}-R_{i})^{-1};
\end{align*}
The computational cost scales as $\sim N_{x}N_{y}^{3}$, so we can take $N_{x}=400$ and the finite-size effect due to the open boundary condition is negligible.

\bibliography{/Dropbox/ResearchNotes/Bibtex/library,Books}
\end{document}